\documentclass[11pt,twoside]{article}

\usepackage{asp2004}
\usepackage{epsf}
\usepackage{lscape}
\usepackage{graphicx}

\begin{document}
\title{The Planets-Capture Model of V838 Monocerotis}   %%% Fill in title
\author{Alon Retter}   %%% Fill in author names
\affil{Department of Astronomy and Astrophysics, Penn State University, 
525 Davey Lab, University Park, PA, 16802-6305, USA;
retter@astro.psu.edu}
\author{Bing Zhang}
\affil{Department of Physics, University of Nevada, Las Vegas, 4505 South 
Maryland Parkway, Las Vegas, NV 89154-4002, USA;
bzhang@physics.unlv.edu}
\author{Lionel Siess}
\affil{Institut d'Astronomie et d'Astrophysique, Universit\'e Libre de 
Bruxelles, CP 226, 1050 Brussels, Belgium; siess@astro.ulb.ac.be}
\author{Amir Levinson}
\affil{School of Physics and Astronomy, Tel Aviv University, Tel Aviv 69978, Israel; levinson@wise.tau.ac.il}
\author{Ariel Marom}
\affil{Rafael Institute, Israel; arielm@rafael.co.il}

\begin{abstract} %%% Abstract to run on from here.
The planets capture model for the eruption of V838 Mon is discussed. 
We used three methods to estimate the location where the planets were 
consumed. There is a nice consistency for the results of the three 
different methods, and we find that the typical stopping / slowing 
radius for the planets is about 1R$_\odot$. The three peaks in the 
optical light curve of V838 Mon are either explained by the swallowing 
of three planets at different radii or by three steps in the slowing 
down process of a single planet. We discuss the other models offered 
for the outburst of V838 Mon, and conclude that the binary merger model 
and the planet/s scenario seem to be the most promising. These two 
models have several similarities, and the main differences are the 
stellar evolutionary stage, and the mass of the accreted material. We 
show that the energy emitted in the V838 Mon event is consistent with 
the planets scenario. We suggest a few explanations for the trigger for
the outburst and for the double structure of the optical peaks in the 
light curve of V838 Mon.
\end{abstract}

\section{Introduction}

V838 Mon had an extraordinary multi-stage outburst during the beginning 
of 2002. Imaging revealed the presence of a spectacular light echo 
around this object (Bond et al. 2003). The amplitude of the outburst 
in the optical band was about 9.5 mag. The post-outburst spectroscopic 
observations of V838 Mon showed that it was very red throughout the 
eruption and long after it ended (Munari et al. 2002; Banerjee \& 
Ashok 2002; Kimeswenger et al. 2002; Evans et al. 2003; Kaminsky \& 
Pavlenko 2005; Tylenda 2005). This is inconsistent with an exposed 
hot white dwarf in novae. 

Evans et al. (2003) and Retter \& Marom (2003) concluded that 
the progenitor star of V838~Mon probably had a radius of $\sim 8 
R_\odot$, a temperature of $\sim 7,300$ K and a luminosity of $\sim 
100-160 L_\odot$. Tylenda, Soker \& Szczerba (2005b) presented a 
detailed analysis of the progenitor. They argued that V838~Mon is 
likely a young binary system that consists of two $5-10 M_\odot$ B 
stars and that the erupting component is a main-sequence or pre-main 
sequence star. They also estimated for the progenitor a temperature 
of $\sim 4,700-30,000$ K and a luminosity of $\sim 550-5,000 L_\odot$. 
Tylenda (2005) adopted a mass of $\sim 8 M_\odot$ and a radius of 
$\sim 5 R_\odot$ for the progenitor of V838~Mon. There is additional 
supporting evidence that the erupting star belongs to a binary system 
with a hot B secondary star (Munari \& Desidera 2002; Wagner \& 
Starrfield 2002; Munari et al. 2005).

Spectral fitting suggested that V838~Mon had a significant expansion 
from a few hundreds to several thousands stellar radii in a couple of 
months during the outburst (Soker \& Tylenda 2003; Retter \& Marom 
2003; Tylenda 2005; Rushton et al. 2005). Interferometric observations 
at the end of 2004 with the Palomar Testbed Interferometer confirmed 
the huge radius of the post-outburst star with an estimate of 
$1,570 \pm 400 R_\odot$ and suggested some asymmetric structure (Lane 
et al. 2005). There are only very rough estimates on the mass of the 
ejecta (Rushton et al. 2003; Lynch et al. 2004; Tylenda 2005).

\subsection{Models for the outburst}

Soon after its outburst, V838~Mon was recognized as the prototype 
of a new class of stars (Munari et al. 2002; Bond et al. 2003), 
which currently consists of three objects: M31RV (Red Variable in M31 
in 1988; Rich et al. 1989; Mould et al. 1990; Bryan \& Royer 1992), 
V4332~Sgr (Luminous Variable in Sgr, 1994; Martini et al. 1999), and 
V838~Mon (Peculiar Red Variable in 2002), plus three candidates -- 
CK~Vul, which was identified with an object that had a nova-like event 
in the year 1670 (Shara \& Moffat 1982; Shara, Moffat \& Webbink 1985; 
Kato 2003; Retter \& Marom 2003), V1148 Sgr, which had a nova outburst 
in 1943 and was reported to have a late type spectrum (Mayall 1949; 
Bond \& Siegel 2006), and the peculiar variable in Crux that erupted 
in 2003 (Della Valle et al. 2003). 

So far, seven explanations for the eruption of these objects have been 
supplied. The first invokes a nova outburst from a compact object, 
which is embedded inside a common red giant envelope (Mould et al. 
1990). In the second model, an atypical nova explosion on the surface 
of a cold white dwarf was suggested (Iben \& Tutukov 1992; Boschi \& 
Munari 2004). Soker \& Tylenda (2003) proposed a scenario in which a 
main sequence star merged with a low-mass star. This model was lately 
revised by Tylenda \& Soker (2006), and summarized by Soker \& Tylenda
(2006). Van-Loon et al. (2004) argued that the eruption was a thermal 
pulse of an AGB star. Munari et al. (2005) explained the outburst of 
V838~Mon by a shell thermonuclear event in the outer envelope of an 
extremely massive (M $\sim 65 M_\odot$) B star. Lawlor (2005, 2006) 
proposed another mechanism for the eruption of V838~Mon. He invoked 
the born-again phenomenon to explain the first peak in the light curve 
and altered the model by adding accretion from a secondary main-sequence 
star in close orbit to explain the second peak in the optical light curve 
of V838~Mon.

A promising model for the peculiar eruption of V838~Mon was suggested 
by Retter \& Marom (2003) and was further developed by Retter et al. 
(2006). This paper summarizes this model for V838 Mon and similar 
objects.

The peculiar and enigmatic outburst of V838 Mon led to a specific 
meeting dedicated to this phenomenon that was held in La Palma, Spain 
on 2006 May, in which the first author of this paper presented the 
planets-swallowing model of V838 Mon. The most important result that 
was presented in the conference is probably two new very reliable 
distance estimates that are consistent with a distance of about 6 kpc 
to V838 Mon (Sparks 2006; Afsar \& Bond 2006). This is somewhat smaller 
than what was previously believed (e.g., Bond et al. 2003), and thus 
it has some impact on the energy emitted in the outburst. 

\section{The planets capture model of V838 Mon}

Retter \& Marom (2003)  showed that the three peaks in the optical 
light curve of V838 Mon have a similar double-shaped structure and 
interpreted them as the devouring of three Jupiter-like massive 
planets by an expanding host star that leaves the main sequence. They 
proposed that it is either a red giant branch (RGB) or an AGB star. 
The planets-swallowing scenario had been analyzed in detail by Siess 
\& Livio (1999a, b).

Retter \& Marom (2003) calculated that the gravitational energy 
released by a Jupiter-like planet that reaches a distance of one
solar radius from the center of a solar-like parent star is sufficient 
to explain the observed eruption. In addition, they found that the 
time scales of the outburst of V838~Mon could be explained by 
this process. Retter \& Marom (2003), therefore, argued that the 
planets-devouring model is generally consistent with the observed 
properties of this object, including its possible binary nature 
mentioned above. This is since planets have been observed in binary 
systems (e.g., Marcy et al. 2005; Mugrauer et al. 2005; Schneider 
2006). 

It was found that the progenitor of V838~Mon is very likely a B star
(Section 1). The planets-swallowing scenario is consistent with a 
B-type progenitor as well. The initial slow expansion of the parent 
star may occur as a result of the natural stellar evolution after 
leaving the main sequence. 

\subsection{Where are the planets consumed?}

Within the planets-devouring model for V838~Mon, Retter et al. (2006) 
estimated the distance from the center of the host star where the 
swallowing process takes place. They used three different methods in 
their calculations: 
(1) checking the energy budget by comparing the observed luminosity 
with the gravitational energy of a Jupiter-like planet that falls 
towards the center of its host star. 
(2) by assuming that a stellar envelope mass of the order of the 
planetary mass is required to stop it or slow it down significantly. 
The resulting timescale was compared with the rise times of the 
peaks in the optical light curve. This method led to the conclusion 
that the critical stellar density, where most of the planetary energy 
is released, is of the order of $\sim 10^{-3}$ gr cm$^{-3}$. From 
density profiles of B stars, the consumption radius was then 
calculated.
(3) using the Roche Lobe geometry. The planet will overflows its Roche 
Lobe only very close to the center of its parent star.

The results from the three methods were consistent with each other 
and suggested that the typical stopping / slowing radius is about 
1$R_\odot$. A careful inspection of the stellar profiles presented in 
fig. 2 in Retter et al. (2006) yielded another insight to this process.
The observed expansion of V838 Mon changed the density profile, and 
therefore, the critical density shifted deeper into the star and closer to 
its core. Thus, the first planet was stopped far away from the stellar
core, while the two other planets had to go deeper. This can explain 
the observed fact that the first peak in the optical light curve is the 
weakest among the three peaks. 

Retter et al. (2006) also suggested an alternative version to the 
three-planets model. They proposed to explain each peak in the optical 
light curve of V838 Mon by a single step in the falling process of a 
single planet. This idea can explain the similar duration of the three 
events.

\section{The signature of planets in stellar envelopes}

Retter et al. (2006) asked the question what would happen to planets
that have recently entered the envelope of their host stars and started
their fall towards its core. They concluded that the planets may show 
quasi-periodic oscillations at the start of this process. When they 
penetrate deeper into the stellar envelope, the amplitude of the 
variations becomes larger, but they are smeared over a longer interval 
of time, and deep inside the parent star the opacity is too large for 
the oscillations to be seen from outside the star. Thus, the falling 
planets can only be detected at the start of the process and at the end 
when they presumably cause an eruption event. This process can explain 
the observed slow rise in brightness in V4332 Sgr before outburst
(Kimeswenger 2006). In addition, Retter (2005, 2006) proposed to
explain the long secondary periods, which are observed in red giant 
stars, and whose origin is still unclear (Wood, Olivier \& Kawaler 
2004), by planets.

\section{The trigger for the outburst}

The values that Retter et al. (2006) derived for the location of the 
accretion process compare very well with the numbers estimated from the 
Virial temperature by Siess \& Livio (1999a). At such a close proximity 
to the stellar core, the temperature of the stellar envelope exceeds 
10$^6$ K. Therefore, the eruption may be triggered by extra energy 
received from the nuclear burning of deuterium brought by the falling 
planets. Another option is that the outburst occurred once the planet 
reached the critical stellar density, which is required to significantly 
slow it down. Hitting denser material causes higher energy release and 
increasing radial acceleration component. At a density of $\rho \sim 
10^{-3}$ gr cm$^{-3}$ and a distance of a few solar radii from the 
stellar core, the opacity, $\kappa$, becomes larger than one. Therefore, 
the trigger for the event could be when the luminosity released by the 
planet is larger than the local Eddington limit. Alternatively, the 
outburst could be triggered by a sudden inward fall of the first planet, 
maybe because of some kind of tidal instability, perhaps due to the 
proximity of the three planets to the parent star and / or to each other, 
or maybe because of eccentric orbits or due to some gravitational 
influence by the secondary star. The consumption of the inner planet and 
the subsequent expansion of the host star led to the engulfment and the 
swallowing of the two other planets. This idea may supply a simple 
solution to the question `how three planets in close orbits around their 
host star can be stable for a long interval of time?', by speculating 
that they were actually unstable. A different idea was presented in 
Section 2.1.

\section{The double structure of the peaks}

Retter \& Marom (2003) showed that the three peaks in the optical 
light curve of V838 Mon have a similar double-shaped structure 
(see their fig.~1), and each of the three peaks is accompanied by a 
shallower peak a few days later. If the trigger for the outburst in 
V838 Mon is when the planet reached the critical stellar density 
(previous section), we may suggest that the three peaks are explained 
by super-Eddington events, when photons are radiated away, while the 
secondary flares can be understood by the material that is ejected 
away at lower velocities and is seen once the opacity is lower than a 
certain limit. This seems like a simple explanation for the time lags 
between the primary and secondary peaks. According to this idea, the 
observed fact that the intervals between the primary and secondary
peaks get longer reflects the stellar expansion.

\section{The rate of planet-swallowing events}

Retter et al. (2006) estimated the rate of V838~Mon-like outbursts within 
the planets-capture model for this phenomenon. They assumed that this is 
a natural step in the stellar evolution and that no unique trigger
mechanism is required for this process. They first started with 
solar-like stars. The number of stars in the Milky Way is about 
10$^{11}$. The age of a $1 M_\odot$ expanding RGB or an AGB star 
(there is a small difference of $\sim 10^{8}$ years between the two 
phases) is about 1.2 $\times 10^{10}$ years (Sackmann, Boothroyd \& 
Kraemer 1993). Thus they obtained a number to age ratio of about 8 per 
year for these stars. The number of B type stars with masses of 
$\sim 5-10 M_{\odot}$ (see Section 1) in our galaxy can be estimated 
as about 1\% of the whole population from the initial mass function 
(e.g. Lucatello et al. 2005). Their evolution is, however, much 
faster than solar-like stars, and their age on the main sequence is 
estimated as about $2-9 \times 10^{7}$ years (Siess 2006). Therefore, 
about $10-50$ massive stars in the Milky Way leave the main sequence 
every year. 

The estimate of the frequency of V838-like outbursts in our galaxy
should take into account the ratio of stars with Jupiter-like planets
in close orbits. Marcy et al. (2005) concluded that about 12\% of FGK 
stars have Jupiter-like planets. Assuming that about 5\% of all stars 
host planets at the relevant range of masses and separations and devour 
them, we thus expect about 0.4 such events per year in our galaxy for 
solar-like stars and $\sim 0.5-2.5$ outbursts in massive stars. 

Many V838~Mon-like eruptions are probably missed. This effect can 
be accounted for by a comparison with nova outbursts because the 
observational bias for these two types of events is similar. About 
$5-10$ novae are detected in our galaxy each year while estimates for 
the actual occurrence number of these eruptions range between 11 and 
260 (Shafter 1997). Adopting a reasonable value of 50 galactic novae 
per year, we estimate that a single V838~Mon-like event should be 
detected every $\sim 2-10$ years in all stars. These values are in 
agreement with the current three members and one candidate in this 
group that erupted in the past 20 years (Section~1.1). Note that the 
wealth of poorly studied novae may hide more V838~Mon-like systems. 
The number of galactic novae that are discovered every year is rising 
fast thanks to many new variability surveys. Therefore, we should 
expect an increase in the frequency of the detection of V838~Mon-like 
events as well.

\section{A Comparison between the different models}

So far seven models have been suggested for the new phenomenon, which 
is defined by V838~Mon (Section 1.1). The binary merger and the 
planet/s-swallowing models seem to have two main advantages over the 
other models. The first is that they can explain an outburst in three 
stages as the optical light curve indicates. In addition, both can 
invoke different types of stars, which is consistent with the 
observations that suggest that a B star responsible for the eruption 
of V838 Mon, while the progenitors of M31RV and V4332 Sgr seem to be 
red giants (Tylenda et al. 2005a; Bond \& Siegel 2006). More arguments 
against the other models can be found in Tylenda \& Soker (2006) and 
Retter et al. (2006).

Tylenda \& Soker (2006) and Soker \& Tylenda (2006) argued that only a 
merger model fits all observed features of the V838 Mon-like stars. 
In their scenario, a low mass ($M \sim 0.1-0.3 M_\odot$) star merged 
with the massive B star, and a third unseen star was probably ejected 
away from the system. This would mean that V838 Mon is a rare quadrapole 
system, because of the massive B companion (Section 1).

Tylenda \& Soker (2006) and Soker \& Tylenda (2006) claimed that the 
energy released by a planet that falls onto a massive star is not 
sufficient to explain the observed eruption. However, as noted by 
Retter et al. (2006) the difference in mass between a low mass stellar 
companion and 1--3 massive Jupiter-like planets is only a factor of 
3--10. In their calculations, Tylenda \& Soker (2006) assumed that the 
falling planet reaches a final distance of about $5 R_\odot$ from the 
center of its $\sim 8 M_\odot$ host star. However, Retter et al. (2006) 
estimated that the consumption occurs much deeper, at a radius of $\sim 1 
R_\odot$. Thus, the energy released by the planet could easily be about 
five times larger than the estimates of Tylenda \& Soker and even higher 
if the planet gets closer to the core of its host star, if it accretes 
some matter during the fall, or if the stellar mass is larger than 
$8 M_\odot$. Therefore, it seems that the energy release by the 
swallowed planets can account for the observed outburst of V838~Mon. 
In addition, the new distance estimate of 6 kpc instead of the previous 
8 kpc (Sparks 2006; Afsar \& Bond 2006) decreases the observed energy 
by a factor of about 2. We also note that 90 percent of the energy 
emitted in the outburst of V838 Mon, estimated by Tylenda \& Soker 
(2006), comes from their estimate for the ejecta mass of $\sim 0.2 
M_\odot$. This is a very unreliable estimate, that could easily be 20 
times smaller. It is enough to point out that most estimates for the 
ejecta mass assumes spherical symmetry, while the observations suggest 
a clear asymmetry (Lane et al. 2005; Wisniewski 2006). We note that 
asymmetric ejection of material is likely to occur both in the binary 
merger and planets-capture models, where a clear preference in the 
orbital plane should take place. Finally, we may speculate that since 
the B star in V838 Mon is very massive, it may have unusually massive 
planets.

An interesting point is that for V4332 Sgr, Tylenda et al. (2005a) and 
Tylenda \& Soker (2006) calculated that its outburst can be explained 
by a merger of a solar-like star and a planet. What is a merger of a 
solar-like star and a planet, if not our scenario?

In summary, we think that only two models among the many offered so far 
for the V838~Mon phenomenon are consistent with the observations. These 
are the binary merger scenario and the planet(s)-swallowing model. These 
ideas are very similar because both invoke the accretion of a secondary 
mass as an explanation for the eruption. Two significant differences 
between the models are the energetics involved and the evolutionary status 
of the donor. The issue of energetics may be answered in the future with 
better modelling and / or observations.

\section{How to distinguish between the models}

As note above the binary merger and the planet(s)-capture models have 
many similarities. How can we distinguish between the two models? It 
is clear that good estimates for the ejecta mass are required. Large 
values would add support to the binary merger model, while low values 
will indicate that the planet(s) scenario is preferred. Other ways of 
obtaining this task are to examine carefully the other members in the 
V838 Mon group, and to find new similar stars. According to the binary 
merger model, the V838 Mon phenomenon should happen among all kinds 
of stars. The planet(s)-swallowing scenario is preferred in old stars, 
that have left the main-sequence -- red giants and asymptotic giant 
branch stars. Note that V838 Mon could be a relatively old B star, 
especially if the dust, which is responsible to the illumination of the 
light echo, originated in previous mass loss episodes in its past. As a
massive star, V838 Mon should have evolved very rapidly.

As a final side note, we comment that the planet capture model is  
consistent with a few of the other models offered for V838 Mon. It is 
possible that V838 Mon swallowed its planets during the large expansion 
in the thermal pulses phase of an asymptotic giant branch star. 
According to the calculations of Lawlor (2005, 2006) accretion of a mass
of the order of Jupiter-like planet is enough to cause the observed 
eruption. Lawlor (2005, 2006) proposed that the accretion comes from a
secondary star, but a planet seems like a nice alternative.

%\subsection{}   %%% Second level section head (remove "%" symbol)
%\subsubsection{}   %%% Lowest level section head (remove "%" symbol)
%\section*{}    %%% Unnumbered top level section head (remove "%" symbol)
%\subsection*{}   %%% Unnumbered second level section head (remove "%" symbol)

\acknowledgements %%% Text of acknowledgements runs on after this command.
This work was partially supported by a research associate fellowship 
from Penn State University. AR is greatful for the AAS for a generous 
international travel grant and for the conference organizers for the 
invitation for the meeting and for some financial support. LS is a FNRS 
Research Associate.

%%% THE BIBLIOGRAPHY
%%%
%%% CONSULT SECTION 4 OF "INSTRUCTIONS FOR AUTHORS" FOR HOW TO USE NATBIB.
%%% PLEASE USE THE "THEBIBLIOGRAPY" ENVIRONMENT
%%%

%\section{Discussion}

\question{A. Evans} One of your predictions was ``infrared emission.''
Can you be more specific?

\answer{A. Retter} This is a prediction by Siess \& Livio (1999a,b).
Check these papers for further details. The idea is basically that 
infrared emission will be a result of mass loss.

\question{R. Hirschi} What speeds up the evolution to the red giant stage
in the planets-capture scenario?

\answer{A. Retter} The release of gravitational energy by the falling 
planet that causes an expansion to large radii.

\question{V.P. Goraskij} We observed planets captured during the main 
outburst when the star was bright, but later, the expansion of the star
continued, and it became faint. In such a condition, the engulfing and 
swallowing of other planets should occur. Why didn't we observe such 
events in the L-supergiant stage?

\answer{A. Retter} Do you want more than three massive Jupiter-like 
planets? It is clear that the falling process of distant planets will 
take longer. 

\question{S. Kimeswenger} As long as the planet is on the main sequence,
the planet will get a lot of irradiation -- why is it not evaporated at 
that time already?

\answer{A. Retter} We think that this will happen for planets with 
Earth-like masses, and that massive Jupiter-like planets should survive 
this radiation.

% edit and fill the lines below with the questions and answers, or
% comment all lines except \end{document}

% \question{SURNAME--QUESTIONER} question text

% \answer{SURNAME--ANSWERER} answer text

\end{document}